\documentclass[a4paper,fleqn,usenatbib]{mnras} 

\usepackage{mathptmx}

\usepackage[T1]{fontenc}
\usepackage{ae,aecompl}

\usepackage{graphicx}	
\usepackage{amsmath}	
\usepackage{amssymb}	

\usepackage{aasmacros}
\usepackage{units}
\usepackage{xspace}

\newcommand{\Msun}{\mathrm{M_{\sun}}}

\newcommand{\kms}{\ensuremath{\rmn{km\,s^{-1}}}}

\newcommand{\Mh}{\ensuremath{M_{\rmn{h}}}\xspace}%
\newcommand{\kpeak}{\ensuremath{k_{\rmn{peak}}}\xspace}%
\newcommand{\NMSM}{$\nu$MSM\xspace}
\def\gsim{ \lower .75ex \hbox{$\sim$} \llap{\raise .27ex \hbox{$>$}} }
\def\lsim{ \lower .75ex \hbox{$\sim$} \llap{\raise .27ex \hbox{$<$}} }

\def\Ms{\ensuremath{M_{\rmn{s}}}\xspace}

\title[Satellites in the \NMSM]{Satellite galaxies in semi-analytic models of
  galaxy formation with sterile neutrino dark matter}
\author[M. R. Lovell et al.]{Mark R. Lovell$^{1,2}$\thanks{E-mail:
    m.r.lovell@uva.nl}, Sownak Bose$^{3}$, Alexey Boyarsky$^{2}$, Shaun Cole$^{3}$, \newauthor Carlos S. Frenk$^{3}$, Violeta Gonzalez-Perez$^{3,4}$,  Rachel Kennedy$^{3}$,  Oleg Ruchayskiy$^{5,6}$ \newauthor and Alex Smith$^{3}$\\
    $^{1}$GRAPPA, Universiteit van Amsterdam, Science Park 904, NL-1098 XH
Amsterdam, The Netherlands\\
$^{2}$Instituut-Lorentz for Theoretical Physics, Niels Bohrweg 2, NL-2333 CA Leiden, The Netherlands\\
$^{3}$Institute for Computational Cosmology, Durham University, South Road, Durham, DH1 3LE, UK\\
$^{4}$Institute of Cosmology and Gravitation, University of Portsmouth, Dennis Sciama Building, Portsmouth PO1 3FX, UK\\
$^{5}$Discovery Center, Niels Bohr Institute, Blegdamsvej 17, DK-2100, Copenhagen, Denmark\\
$^{6}$\'Ecole Polytechnique F\'ed\'erale de Lausanne, FSB/ITP/LPPC, BSP 720, CH-1015, Lausanne, Switzerland\\}

\date{Accepted *** Received ***; in original
  form ***} 
\pubyear{2015}

\begin{document}
\label{firstpage}
\pagerange{\pageref{firstpage}--\pageref{lastpage}} 

\maketitle

\begin{abstract}
  The sterile neutrino is a viable dark matter candidate that can be produced in the early Universe via non-equilibrium processes, and would therefore possess a highly non-thermal spectrum of primordial velocities.
  In this paper we analyse the process of structure formation with this class of dark matter particles.
  To this end  we construct primordial dark matter power spectra as a function of the lepton asymmetry, $L_6$, that is present in the primordial plasma and leads to resonant sterile neutrino production. 
  We compare these power spectra with those of thermally produced dark matter particles and show that resonantly produced sterile neutrinos are much colder than their thermal relic counterparts. 
  We also demonstrate that the shape of these power spectra is not determined by the free-streaming scale alone.
  We then use the power spectra as an input for semi-analytic models of galaxy formation in order to predict the number of luminous satellite galaxies in a Milky Way-like halo.
  By assuming that the mass of the Milky Way halo must be no more than $2\times10^{12}\Msun$ (the adopted upper bound based on current astronomical observations) we are able to constrain the value of $L_6$  for $\Ms\le 8$~keV.
  We also show that the range of $L_6$ that is in best agreement with the 3.5~keV line (if produced by decays of 7~keV sterile neutrino) requires that the Milky Way halo has a mass no smaller  than $1.5\times10^{12}\Msun$.  
 Finally, we compare the power spectra obtained by direct integration of the Boltzmann equations for a non-resonantly produced sterile neutrino with the fitting formula of Viel~et~al. and find that the latter significantly underestimates the power amplitude on scales relevant to satellite galaxies.
\end{abstract}

\begin{keywords}
  dark matter 
\end{keywords}

  \section{Introduction}
  \label{intro}

  The identity and properties of  dark matter remain among the most
  pressing questions in physics. For some 30 yr it has been
  understood that the dark matter particle should be kinematically
  cold \citep{Davis85} and non-baryonic in order to satisfy
  nucleosynthesis constraints and microwave background temperature anisotropy data \citep{Larson_11,PlanckCP15}. However the precise identity of the dark matter particle remains elusive. 

  Many experiments have been devised to look for hypothetical weakly interacting massive particles (WIMPs) predicted by supersymmetric theories \citep{Ellis_84}, models with extra dimensions~\citep{Servant:2002aq} and other models with new physics at the TeV scale. 
Although the CDMS-Si and DAMA  direct detection experiments have reported signals consistent with WIMP dark matter \citep{Agnese13,DAMA13}, other experiments such as XENON100 and LUX find the same parameter space to be ruled out \citep{XENON2012, LUX2014}. 
Other hints have come from the detection of unexplained astronomical signals which could be provided by annihilating WIMPs, such as $\gamma$-ray lines \citep[e.g.][]{Weniger12,Daylan:2014rsa}, $\gamma$-ray excesses \citep{Hooper11,Calore:2014xka}, or cosmic ray positron excesses \citep{AMS13}. However, many of these signals have since been ruled out \citep{Ackermann13,Boyarsky:2012ca}, or could potentially be explained by other astrophysical sources \citep[e.g.][]{Bartels15}. Thus, after over a decade of increasingly sensitive experiments, a conclusive and consistent signal has not yet emerged.

The weak interaction of GeV--TeV dark matter candidates with ordinary matter requires a new symmetry to stabilize these particles from fast decays. 
If one does not assume such a symmetry, the dark matter candidate should be \emph{super-weakly interacting} and/or light.
One such candidate of particular interest is the sterile neutrino \citep{Dodelson94,Shi99,Abazajian:01b,Abazajian:01a,Asaka:05a,Asaka:05b,Asaka:06c}, see~\citet{Boyarsky12} for review.
The existence of sterile neutrinos is motivated by the phenomenon of neutrino oscillations: the change of neutrino flavours could be due to mixing with these new particles. 
However, it has been recognized \citep{Asaka:05a,Boyarsky:06a} that sterile neutrino dark matter particles do not contribute significantly to the neutrino oscillations. 
Therefore, at least three sterile neutrinos are required to explain the neutrino masses and mixings and provide a dark matter candidate.
It turns out \citep{Asaka:05b,Canetti:12,Canetti:12b} that these same sterile neutrinos can be responsible for the generation of the matter-antimatter asymmetry in the early Universe. This three sterile neutrino model, with their masses below the electroweak scale, explains neutrino oscillations, the mechanism of baryogenesis and the dark matter candidate, and has been named the Neutrino Minimal Standard Model or \NMSM (see \citet{Boyarsky09a} for review).

Sterile neutrino dark matter particles in the \NMSM have keV-scale mass and thus have a sufficiently large free-streaming velocity to act as warm dark matter (WDM), see Section~\ref{sec:snm} below. The free-streaming of WDM particles erases perturbations in the early Universe below the \emph{free-streaming horizon} and therefore suppresses the formation of dark matter haloes that could otherwise contain dwarf and satellite galaxies \citep[see e.g.][]{Weinberg:13}.
WDM models may be able therefore to ease tensions between simulated dark matter haloes and satellite galaxies, in both their abundance and structure \citep{Bode01,Polisensky2011,Lovell12,Anderhalden13} ; in this role WDM is an alternative or even a complement to proposed astrophysics-based processes such as reionization \citep{Benson02} and supernova feedback \citep{Pontzen_Governato_11,Zolotov2012, Governato15}. 
  The abundance of satellite galaxies in particular is an attractive property to
  consider \citep{Polisensky2011,Kennedy14,Lovell14, Schneider14b}. \citet{Kennedy14} used a semi-analytic galaxy formation model to constrain the properties of WDM particles.
In this study we extend the analysis of \citet{Kennedy14} to sterile neutrino dark matter in order to obtain constraints on both the mass of the sterile neutrino and an additional particle physics parameter, the lepton asymmetry with which the sterile neutrinos are generated. 

Interest in sterile neutrinos has increased recently since multiple reported detections of an X-ray line from the {\it XMM-Newton}, {\it Chandra}, and {\it Suzaku} X-ray observatories at an energy of 3.5~keV in the Galactic Centre, M31 and numerous galaxy clusters \citep{Boyarsky14a,Boyarsky14b, Bulbul14,Iakubovskyi15}, which could be generated by the decay of a 7~keV sterile neutrino. This hypothesis has been tested with new, deep observations of the Draco dwarf galaxy, however the results are inconclusive \citep{Jeltema15,Ruchayskiy16}. Below we pay special attention to the properties of such 7~keV particles.

This paper is organized as follows. In Section~\ref{sec:snm} we
provide a review of the sterile neutrino model and
discuss its influence on the matter power spectrum. The method for calculating the abundance of satellite galaxies is
described in Section~\ref{sec:gfm} and the results presented in
Section~\ref{sec:res}. We draw conclusions in Section~\ref{sec:con}.

\section{Sterile neutrino dark matter}
\label{sec:snm}

In this section we provide an outline of how the sterile neutrinos are produced, describe their momentum distribution functions and matter power spectra, and discuss the differences with generic thermal relic WDM.

\subsection{Production mechanisms}

Sterile neutrinos were first suggested as a dark matter candidate by \citet{Dodelson94}. Their model considered a single sterile neutrino added to the standard model of particle physics.  This new particle would be produced through mixing with active neutrinos in the early Universe.  Although produced out of equilibrium, its primordial momentum distribution would resemble that of a (rescaled) Fermi-Dirac distribution~\citep{Dolgov:00}. The production rate peaks at a temperature $T_\text{prod}\sim \unit[150]{MeV}(M_{\rmn{s}}/\unit[1]{~keV})^{1/3}$ \citep{Dolgov:00,Asaka:06c} and the average momentum scales as $\langle p\rangle \sim T_\text{prod}$; here \Ms denotes the sterile neutrino mass. The particles are therefore produced relativistically in the range of masses up to $\sim 2$~MeV (i.e. their average momentum, $\langle p\rangle \gsim$ \Ms), and thus the sterile neutrino
particle is a WDM  candidate. 
  
The WDM nature of these particles, together with the fact that sterile neutrinos decay into X-rays on a time-scale much longer than the age of the
Universe~\citep{Abazajian:01b,Dolgov:00}, has enabled very strong constraints to be placed on the particle's properties: low mass sterile neutrinos would free-stream out of large scale ($k>1~h~\rmn{Mpc}^{-1}$) primordial perturbations and prevent galaxies from forming, whilst high mass sterile neutrinos would be readily detectable in X-rays since the particle lifetime, $\tau \propto M_{\rmn{s}}^{-5}$. The combination of these two constraints ultimately excluded the entire available sterile neutrino mass range of interest, and thus ruled out a purely non-resonantly produced sterile neutrino \citep{Seljak:06,Viel:06,Boyarsky:08c}.

However, being produced out of thermal equilibrium, sterile neutrinos are sensitive to the content of primordial plasma. In particular, the production of sterile neutrinos could be enhanced in the presence of a lepton asymmetry, i.e. an overabundance of leptons relative to anti-leptons \citep{Shi99}. In this case, the lepton asymmetry increases the effective sterile neutrino mixing angle, much in the same way as the Mikheyev-Smirnov-Wolfenstein (MSW) effect~\citep[][]{Wolfenstein:1977ue,Mikheev:1986gs} changes the mixing angles of active neutrinos passing through a medium. 
The lepton asymmetry required for this mechanism to work is much greater than the inferred value of the baryon asymmetry of the Universe, and should exist in the primordial plasma \emph{after} the electroweak transition.

 A sufficiently large lepton asymmetry appears naturally 
for a large part of the parameter space of the \NMSM model in the temperature range 10-100 GeV \citep{Shaposhnikov:08a,Canetti:12,Canetti:12b}, being generated by two other, unstable sterile neutrino species at the GeV scale, and leads to the resonant production of dark matter particles \citep{Laine08}.  The resonant enhancement applies to momenta below some threshold, so the dark matter can be cooler than in the non-resonant case and thus no longer in tension with structure formation bounds~\citep{Boyarsky:08d}.

\begin{figure}
    \includegraphics[scale=0.34]{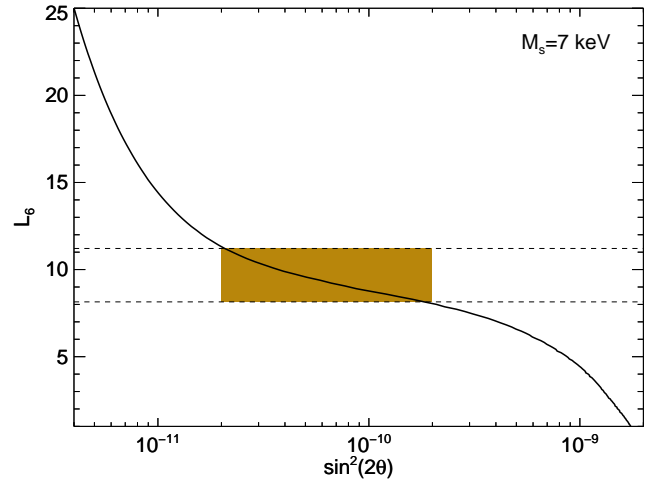}  
    \caption{$L_6$ as a function of the mixing angle,
    $\sin^{2}(2\theta)$, for the 7~keV sterile neutrino,  such that the sterile neutrino abundance is equal to the dark matter abundance. The brown
    region delineates the mixing angle range inferred for the 3.5~keV
    line by \citet{Boyarsky14a}. Further examples may be found in
    \citet{Boyarsky09a}; see also \citet{Abazajian14}.}
    \label{TTvsL6}
  \end{figure}

The lightest of the three sterile neutrinos in the \NMSM is a dark matter candidate whose properties are determined by three parameters: its mass, \Ms\footnote{In some particle physics studies, this mass is denoted $M_1$ so as to be distinguished from the two GeV-scale unstable sterile neutrino masses, $M_2$ and $M_3$. }, the mixing angle, $\theta_1$ -- which together with \Ms determines the X-ray decay rate flux, $F\propto \sin^{2}(2\theta_{1})\Ms^{5}$ \citep{Pal:81,Barger:95} -- and the lepton asymmetry, which we denote as $L_6$.  Here we define $L_6$ as $10^6$ times the difference in electron neutrino and anti-electron neutrino abundance divided by the entropy density, i.e.  $L_6 \equiv 10^6 \frac{n_{\nu_\rmn{e}} - n_{\bar\nu_\rmn{e}}}s$; for an alternative parametrization see \citet{Abazajian14}. The requirement that the correct cosmic dark matter density parameter be obtained has the effect that setting the value of two of the parameters determines uniquely the value of the third. 
For a given mass, as lepton asymmetry increases the mixing angle must decrease. We show the relationship between these two parameters for the 7~keV sterile neutrino in Fig.~\ref{TTvsL6}, highlighting the range of $\theta_1$ -- and thus $L_6$ -- that is consistent with the 3.5~keV line. For the remainder of this study we retain $L_6$ as the free parameter rather than the mixing angle.

\subsection{Primordial velocities of sterile neutrino dark matter} 

The lepton asymmetry has an important effect on the shape of the sterile neutrino momentum distribution. It is parametrized in terms of $L_6$ as defined in the previous subsection; $L_6 = 0$ thus corresponds to the absence of any lepton asymmetry. As explained in \citet{Boyarsky09a} the maximal lepton asymmetry attainable in principle within the \NMSM is   $L_6^{\rm max} = 700$ [for comparison the big bang nucleosynthesis (BBN) bound on primordial lepton asymmetry of~\citet{Serpico:05} would correspond to $L_6^{\rm BBN} \simeq 2500$] therefore we choose to perform the analysis in the range $0 \le L_6 \le 700$.\footnote{The recent work of \citet{Canetti:12,Canetti:12b} has not revealed any combination of the \NMSM parameters that would lead to $L_6$ in excess of $120$. However this is a work in progress and therefore we choose to explore a wider region of the parameters.}

\begin{figure}	
  \includegraphics[width=\linewidth]{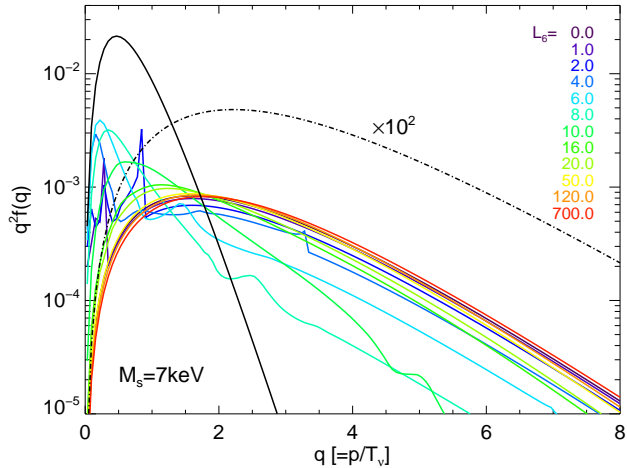} 
  \caption{Sterile neutrino momentum distributions for different values of $L_6=[0,700]$ as indicated by the legend.  Each distribution $f(q)$ is multiplied by the momentum squared, $q^2 \times f(q)$, to reduce the dynamic range. Solid black line -- the Fermi-Dirac distribution of a thermal relic with the mass $M_{\rm th} = 1.4$~keV (the temperature of this distribution is different from $T_\nu$). Dashed black line -- Fermi-Dirac distribution with $T=T_\nu$, multiplied by $10^{-2}$ to fit into the plot. The momenta are plotted for the thermal plasma at a temperature of 1~MeV. The sterile neutrino mass is 7~keV.}
  \label{DFexample}
\end{figure}

For a given value of \Ms we can therefore vary the value of $L_{6}$ and attain a variety of different momentum distributions. As a pertinent example of how the momentum distribution changes for different lepton asymmetry, we generate the  \Ms=7~keV  sterile neutrino momentum distributions for 12 values of $L_6$ using the methods and code of \citet{Laine08} and \citet{Ghiglieri15}, and plot the results in Fig.~\ref{DFexample} (we do not expect the most recent computations by~\citealp{Venumadhav:2015pla} to affect our results).

The behaviour of the momentum distribution is non-monotonic with increasing $L_6$. For $L_6=0$ (non-resonant production, NRP) the distribution is well approximated by a rescaled Fermi-Dirac distribution, at least to first order. As $L_6$ increases, a prominent resonance spike appears at low momentum, along with several additional spikes at higher momenta. The amplitude and position of these resonance peaks increase with lepton asymmetry such that the overall spectrum becomes cooler.\footnote{The precise shape of the distribution function at low momentum is complicated by the effect of the Pauli exclusion principle. Because of it some low $q$ sterile neutrinos may be forced to acquire a higher momentum, an effect known as a `back reaction' \citep{Laine08}. There is still some uncertainty as to which approximations should be used to calculate the contribution of this effect accurately, and this uncertainty translates into a systematic error on the distribution shape. In this study we choose a back-reaction prescription of \citet{Laine08} that gives the coldest distribution functions for given mass and lepton asymmetry.}  Eventually the resonance peak shifts to sufficiently high momentum that the spectrum becomes warmer again, and for $L_6 \gtrsim 50$ the momentum distribution is very similar to that for non-resonant production. This is because the resonance is enhancing production at all momenta.

\begin{figure}
    \includegraphics[width=\linewidth]{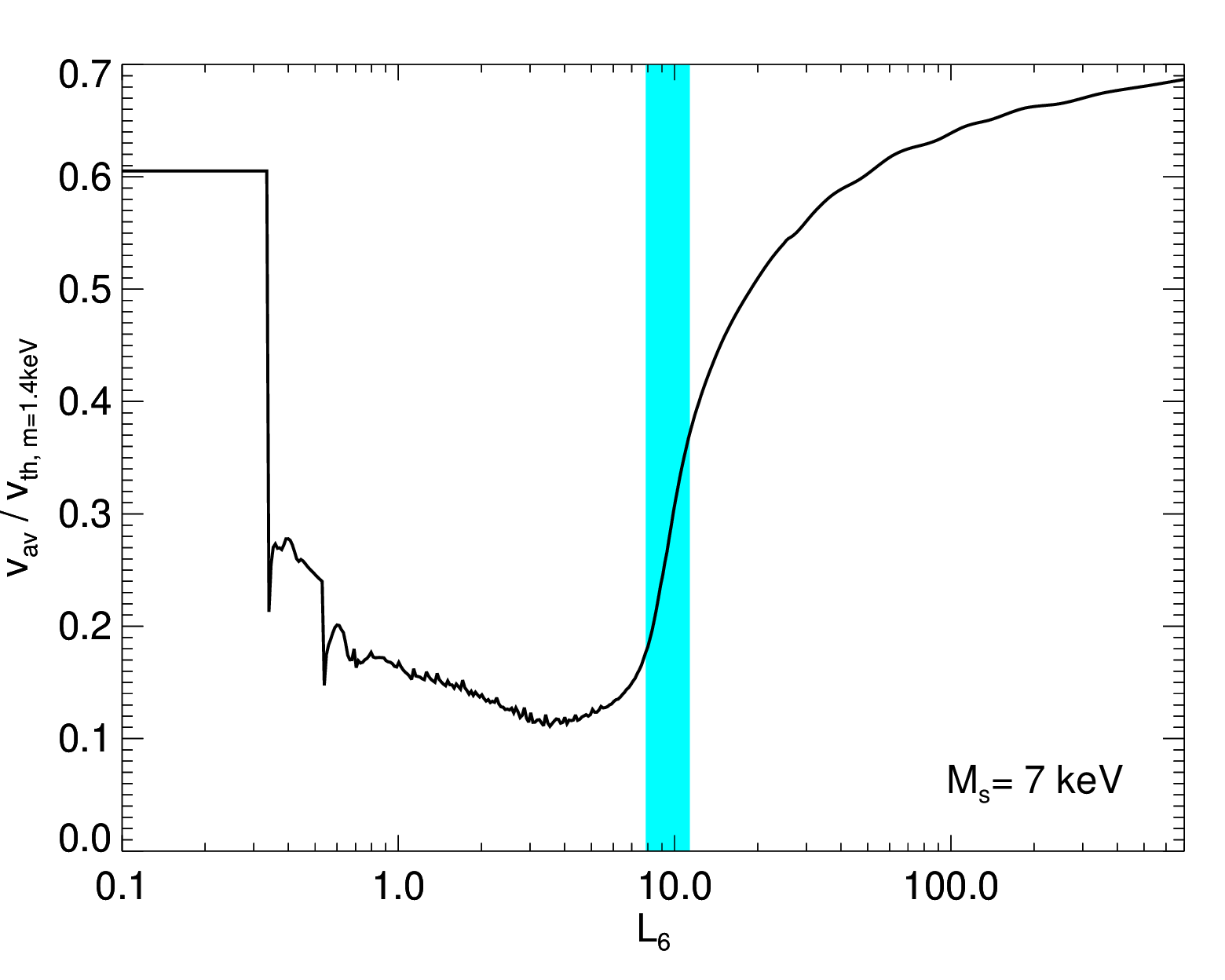}  
    \caption{The average velocity (a proxy for the free-streaming length), at matter-radiation equality for a series of \Ms=7~keV momentum distributions as a function of lepton asymmetry, normalized to the corresponding average velocity of the thermal relic with $M_{\rm th} = 1.4$~keV, $v_{\rm th} = \unit[54]\kms$. The region of $L_6$ that is consistent with the reported 3.5~keV line \protect\citep{Boyarsky14a} is given by the cyan region. Small wiggles on the curve correspond to the contributions of  sharp resonant peaks in the distribution functions, as seen in Fig.\protect~\ref{DFexample}.}
    \label{L6vsQav}
  \end{figure}

 The shape of these momentum distributions is a key ingredient for calculating the free-streaming velocity, the magnitude of which influences the shape of the matter power spectrum. As a proxy for the free-streaming velocity we calculate the average velocity of the momentum distributions at matter-radiation equality; we denote this average velocity by $v_{\rmn{av}}$. We calculate $v_{\rmn{av}}$ for a grid of 500 7~keV momentum distributions, each of which has a value of $L_6$ in the range [0.1, 700], and plot the results in Fig.~\ref{L6vsQav}. We find that for $L_6<0.3$, $v_{\rmn{av}}$ has a constant value of $32~\kms$ (which is about 40 per cent lower than the  
 $v_{\rmn{av}}$ of Fermi-Dirac distribution with the same mass and $T=T_\nu$ or, equivalently, of the thermal relic particle with the mass $M_{\rm th} = 1.4$~keV).  
The average velocity then sharply decreases with $L_6$ as the resonance begins to take effect, and attains a minimum value at $L_6\approx5$ before rising once again to values higher even than that of the non-resonant case. 
Thus, there exists a `sweet spot' for $L_6$ for which the spectrum is maximally cool, either side of which the average free-streaming velocity rises rapidly.
  
\subsection{Power spectra of sterile neutrino dark matter}

\begin{figure}
  \includegraphics[scale=0.34]{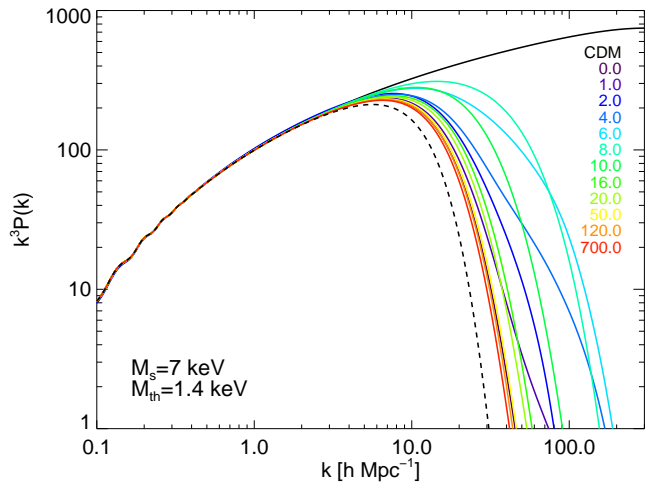}  
  \caption{Matter power spectra generated from the \Ms=7~keV distribution functions shown in Fig.~\ref{DFexample}. The CDM power spectrum is shown as a solid black line. The dashed line corresponds to the power spectrum of a thermal relic of mass 1.4~keV, which is the thermal relic counterpart of the NRP 7~keV sterile neutrino, as derived from equations~\ref{SterTherm} and~\ref{VielEq}. See the legend for correspondence between colour and lepton asymmetry value.}
  \label{MPexample}
\end{figure}

\begin{figure}
   \includegraphics[scale=0.34]{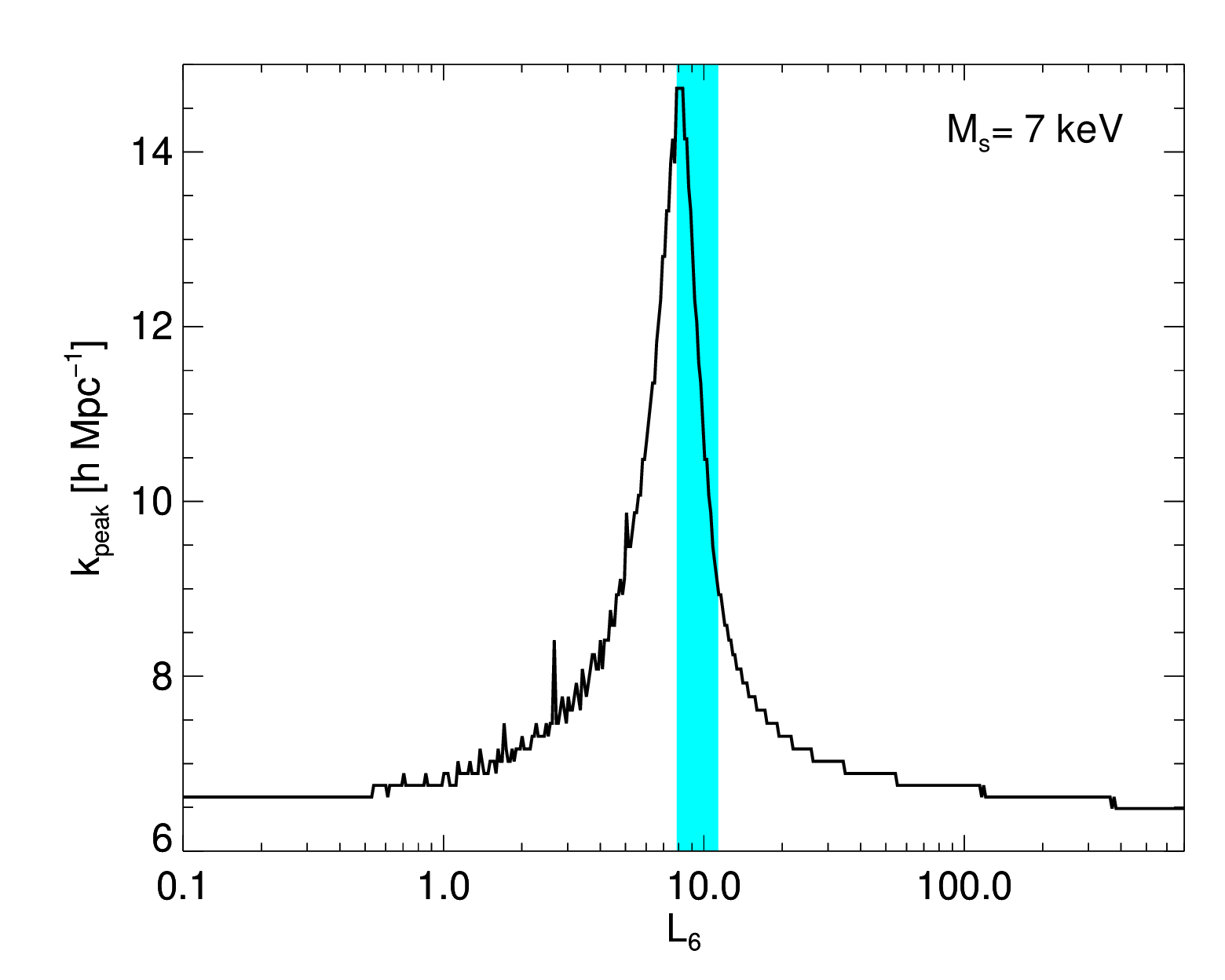}
   \includegraphics[scale=0.34]{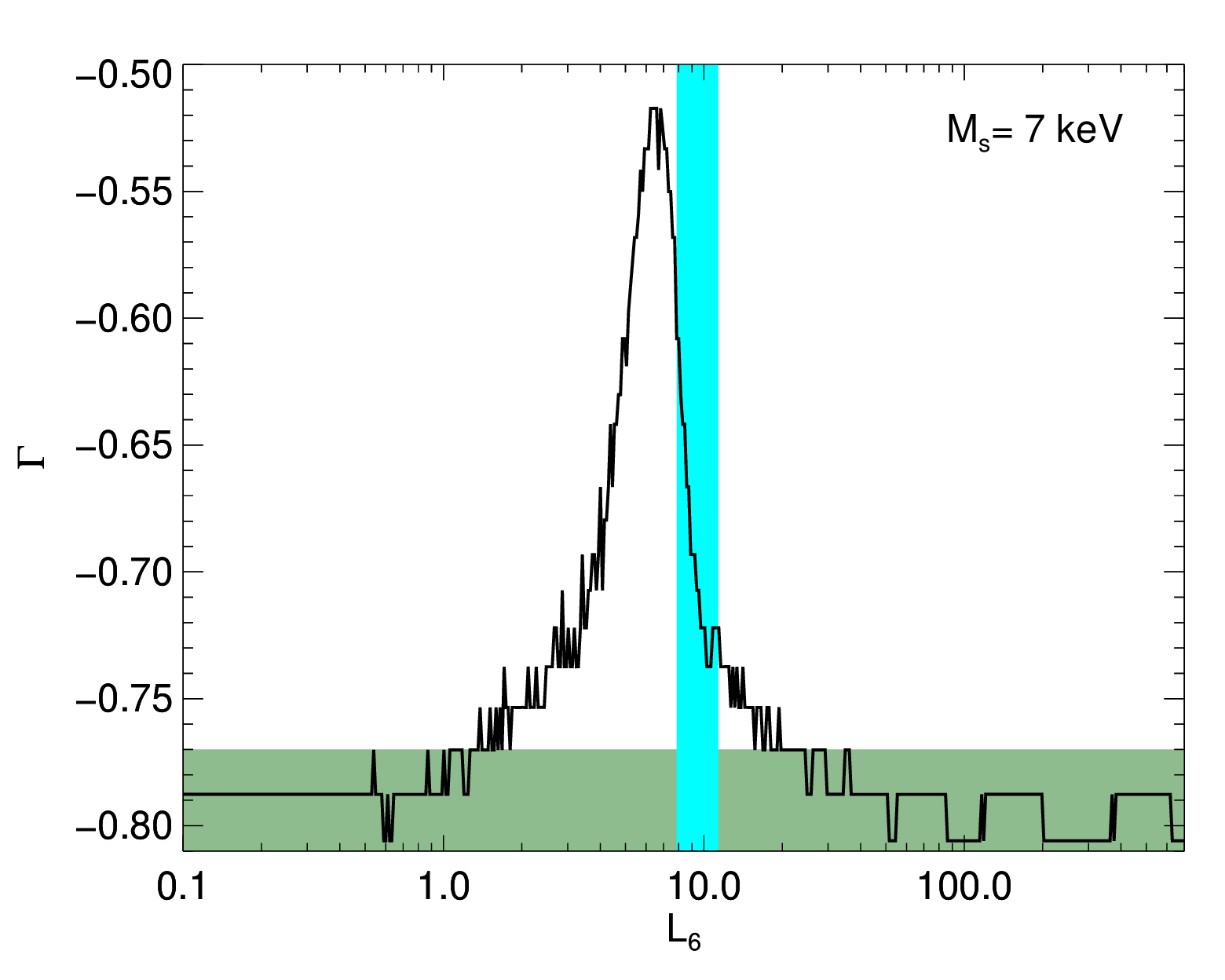}
   \caption{Properties of the \Ms=7~keV matter power spectrum as a function
   of $L_6$. Top:  \kpeak; bottom: 
   logarithmic gradient. The region in $L_6$ consistent with the 3.5~keV
   line is shown in cyan. The dark green region marks the range of $\Gamma$ for V05 thermal relics with the same \kpeak as our 7~keV matter power spectra.}
   \label{L6vsPG}
\end{figure}

The shape of primordial momentum distribution is imprinted in the linear matter power spectrum. We use the distribution functions calculated for Fig.~\ref{DFexample} as inputs to a modified version of the Boltzmann solver \textsc{camb} \citep{Lewis2000} in order to derive the sterile neutrino dimensionless linear matter power spectra -- $\Delta^{2}(k)=k^{3}P(k)$ -- at redshift zero; we assume the \emph{Planck} cosmological parameters (quoted in Section~\ref{sec:gfm}). We plot the results in Fig.~\ref{MPexample}.
As anticipated from the distribution functions, the wavenumber at which the dimensionless matter power spectrum attains its peak amplitude, \kpeak, increases with $L_6$ up to some value and then returns to the original, warmer spectrum. These power spectra each have a pronounced cutoff as is the case for fiducial WDM. However, the power spectra close to the turnover have a shallower slope than for the minimally or maximally resonant sterile neutrino \footnote{This shallow slope is similar to that of a cold + warm dark matter mixture (CWDM)  
  for values of $k$ that are close to the cutoff, and thus CWDM has been used in the past to approximate sterile neutrino matter power spectra for studies of the Lyman-$\alpha$ forest \citep{Boyarsky:08d, Boyarsky:08c}. However, for high resolution simulations and Monte Carlo Markov chain methods this approximation is not appropriate as it
  introduces spurious excess power at small
  scales.}. In order to examine the evolution of the cutoff position and gradient more quantitatively,  in Fig.~\ref{L6vsPG} we display both \kpeak and the turnover gradient for a set of 500
  7~keV  matter power spectra as a function of $L_6$; we define the turnover gradient, $\Gamma$, as the logarithmic gradient between $k_\rmn{peak}$ and $k_{\rmn{half-peak}}$, where the latter is defined as $\Delta^2({k_{\rmn{half-peak}} > \kpeak}) = 0.5\Delta^2(\kpeak)$. 

The value of \kpeak increases by over a factor of 2
between the coolest and warmest models, from $6.6~h~\rmn{Mpc}^{-1}$ at $L_6=0.1$ to a maximum $14.7~h~\rmn{Mpc}^{-1}$ at $L_6=8$, which is a higher value of the lepton asymmetry than that at which $v_{\rmn{av}}$ is minimized. We can therefore conclude that knowledge of the average momentum (and thus free-streaming velocity) is not sufficient to determine the  matter power spectrum and that the precise shape of the momentum distribution function therefore plays a key role (see also~\citealp{Maccio':2012uh} where this has been demonstrated for the cold+warm dark matter models, CWDM). The gradient exhibits similar behaviour: the cutoff is sharp at the largest and smallest values of $L_6$, and is comparable to that of the slope of the WDM transfer function developed by \citet[][hereafter V05]{Viel05}, but becomes much shallower in the middle of the range plotted. The shallowest gradient is attained at $L_6\sim6.3$, thus at slightly lower $L_6$ than is the case for \kpeak. We explore the role of these two parameters, plus $v_{\rmn{av}}$, for predicting the Milky Way satellite galaxy abundance in Appendix~\ref{app:char}.  
  
It is interesting to compare the shapes of sterile neutrino power spectra with the matter power spectra of thermal relic particles. \citet{Colombi96} and \citet{Bode01} showed that an NRP sterile neutrino matter power spectrum, as derived from \citet{Dodelson94}, would have precisely the same shape as a thermally produced particle of some mass. V05 built upon these studies and derived the following relationship between the sterile neutrino and thermal relic masses:
\begin{equation}
 \Ms = 4.43\rmn{~keV}\left(\frac{M_{\rmn{th}}}{1\rmn{~keV}}\right)^{4/3}\left(\frac{0.7^{2}\times0.25}{h^{2}\Omega_{\rmn{DM}}}\right)^{1/3},
  \label{SterTherm}
\end{equation}
where $M_{\rmn{th}}$ is the thermal relic mass, $h$ the Hubble parameter and $\Omega_{\rmn{DM}}$ the dark matter density parameter. Thus, a 7~keV NRP sterile neutrino would lead to the same matter power spectrum as a 1.4~keV thermal relic.  V05 then generated a series of thermal relic matter power spectra using Boltzmann codes such as {\sc camb}, and found that their results could be well fit at large $k$ by the equation $P_{\rmn{WDM}}(k)=T^{2}(k)P_{\rmn{CDM}}$, where the transfer function $T(k)$ has the functional form:
\begin{equation}
  T(k) = [1+(\alpha k)^{2\nu}]^{-5/\nu}.
  \label{VielEq}
\end{equation}
Here $\nu$  is a parameter determined by fitting to the full Boltzmann code transfer function; V05 find a best-fitting value of $\nu=1.12$. The $\alpha$ parameter is a function that depends on the thermal relic mass (see equation~7 of V05). This fitting function and parametrization provide a good fit for $k<5h~\rmn{Mpc}^{-1}$.
   
   We now analyse the correspondence between this fitting function and our matter power spectra. Since the work of \citet{Dodelson94}, further developments in the calculation of sterile neutrino properties have broken the precise link between thermal relic mass and sterile neutrino mass given in equation~\eqref{SterTherm} \citep{Abazajian:01a,Abazajian:02,Dolgov:00,Asaka:06b,Asaka:06c}. To make this point explicit, we plot the V05 curve for a 1.4~keV thermal relic in Fig.~\ref{MPexample}. We find that the agreement between the `true' NRP sterile neutrino curve and the derived thermal relic curve is still better than 10 per cent at $k\le5~h~\rmn{Mpc}^{-1}$. However, the thermal relic has considerably less power at larger $k$. The thermal relic curve peaks at a $k=5.5~h~\rmn{Mpc}^{-1}$, 10 per cent below the sterile neutrino peak. It also exhibits a much more rapid cutoff than the sterile neutrino, such that by $k\approx12~h~\rmn{Mpc}^{-1}$ the amplitude of the thermal relic power spectrum is only half that of the sterile neutrino spectrum. 
   
   In order to be check our calculation, we generated thermal relic matter power spectra using our modified version of {\sc camb}. The agreement around the peak in $\Delta^2$ between the thermal relic Boltzmann calculations and the V05 fit is at the per cent level. We also note that recently \citet{Paduroiu15} argued that previous studies had underestimated the free-streaming length of thermal relics, which would further increase the tension with sterile neutrino properties.
   
    The scales around the cutoff play an important role in structure formation, and thus this discrepancy between thermal relics and sterile neutrinos is important for this study. Since most simulation-based studies of sterile neutrinos and generic WDM have been performed using the V05 transfer function, care should be taken when comparing our results with previous work. For the remainder of this paper we use the {\sc camb}-derived sterile neutrino transfer functions except where stated otherwise.

  \section{Methods}
  \label{sec:gfm}

  In order to calculate the satellite galaxy abundance, and
  consequently place limits on the sterile neutrino mass, we employ the same
  methodology as \citet{Kennedy14}, only now applied to sterile neutrinos rather than thermal relics. We use our sterile neutrino power spectra, which are now functions of the sterile neutrino mass and lepton asymmetry as shown in Fig.~\ref{MPexample}, to generate dark matter halo merger trees for a series of host halo masses. These merger trees are combined with a state-of-the-art semi-analytic galaxy formation model to generate populations of satellite galaxies. The number of satellites is then compared to the measured abundance of Milky Way satellites. The number of dark matter subhaloes above some mass, and thus the number of satellites galaxies, is roughly proportional to the mass of the halo \citep[][ and references therein]{Wang12}. If the number of satellites resulting from a given combination of sterile neutrino mass, lepton asymmetry, and host halo mass fails to produce enough satellites to meet the number expected for the Milky Way then that combination is ruled out. We can then use estimates on the mass of the Milky Way halo to constrain the sterile neutrino mass and lepton asymmetry.
    
    Our chosen semi-analytic
  model is the \citet{Gonzalez14} version of the \textsc{galform} model of galaxy formation
  \citep{Cole00,Benson03,Bower06}. This model calculates the evolution of galaxies throughout cosmic history, following the formation of dark matter haloes, the accretion and shock-heating of infalling gas that subsequently cools to form stars and the accretion of satellites on to haloes. It accounts for processes such as the photoionization and the chemical enrichment of gas and stars. Star formation is regulated by feedback from both active galactic nuclei (AGN) accretion and, very importantly on the mass scales of interest here, supernovae. The expulsion of gas from the galaxy due to supernova feedback is modelled assuming that the mass loading is a power-law of the galaxy circular velocity; the power law index, denoted $\alpha_{\rmn{hot}}$, with a fiducial value of 3.2. This value is determined by calibrating to observations of the $z=0$ galaxy luminosity function assuming cold dark matter (CDM). It should therefore be re-calibrated for sterile neutrino models. \citet{Kennedy14} found a value of $\alpha_{\rmn{hot}}=3.0$ (with a slight dependence on the warm dark matter particle mass), and we adopt this value throughout this study except where stated otherwise. A discussion of the re-calibration of the model may be found in Appendix~\ref{AFC}. The fiducial (i.e. CDM) model parameters as a whole are determined by calibrating their values to make reasonable predictions for the evolution of the rest-frame $K$-band and optical luminosity functions simultaneously: a comprehensive discussion can be found in \citet{Gonzalez14}.

  We apply this model to merger
  trees of dark matter haloes generated using the
  extended Press-Schechter (EPS) formalism
  \citep[][]{Press74,Bond91}, and assume the \emph{Planck} cosmological
  parameters: $\Omega_{0}=0.307$, $h = 0.678$, $\Omega_{\Lambda}=0.693$, $\Omega_{\rmn{b}}=0.0483$, $\sigma_{8}=0.823$, and $n_{\rmn{s}}=0.961$.
  
\subsection{EPS implementation}

The rms density fluctuation, $\sigma(M)$, can be written as
  
  \begin{equation}
    \sigma^2(M) = \frac{1}{2\pi^2} \int^{\infty}_0 k^2 P(k) W^2(k;M) \rmn{d}k,
  \end{equation}

  \noindent
  where $W_k(k;M)$ is a window function which, in the standard EPS
  method, is chosen to be a top hat in real space. In Fourier space,
  this window function has the form
  $W(k;M)=3[\sin(kR)-kR\cos(kR)]/(kR)^3$, where the mass, $M$, and
  filtering scale, $R$, are unambiguously related through $M =
  \frac{4}{3} \pi \bar{\rho} R^3$. Decreasing the mass of the window
  function has the effect of reweighting large $k$ modes. This means
  that if the power spectrum has a sharp cutoff, $\sigma(M)$ will
  continue to increase with decreasing $M$, even though no new modes
  enter the filter.
  
   An alternative to the real space top hat is a sharp $k$-space
  filter. With this choice, the flattening of $\sigma(M)$ is set
  entirely by the sharpness of the power spectrum cutoff. However,
  this raises the problem that it is no longer clear how to relate the
  mass to the filtering scale. On dimensional grounds $M \propto
  k_{\rm cut}^{-3}$,  and to maintain the usual relation between mass 
and radius we can write $M_{\rm SK} = \frac{4}{3} \pi
  \bar{\rho} R_{\rm SK}^3$, with $R_{\rm SK}=a/k_{\rm cut}$
  where $a$ is a constant which needs to be determined. By integrating the mean density under the window
  function and setting this equal to the required mass,
  \citet{Lacey93} find a value of $a = (9\pi/2)^{1/3} \approx
  2.42$. \citet{Benson13} and \citet{Schneider13} match their results
  to \emph{N}-body simulations, and find values of $a=2.5$ and $a=2.7$
  respectively.

  The use of a sharp $k$-space filter has been shown to be a
  suitable approach for the V05 transfer function \citep{Benson13}, which has a
  cutoff sharper than many of our models, as shown in
  Section~\ref{sec:snm}. To check that the method is still valid for
  our shallower sterile neutrino dark matter cutoffs it is necessary
  to check the calibration against \emph{N}-body simulations. 
  
  To this end, we identified which of our set of sterile neutrino
  matter power spectra has the shallowest cutoff of the whole set
  -- \Ms = 3 ~keV and $L_6=14$, hereafter M3L14 -- and used this as the
  input transfer function for re-runs of four of the Aquarius Project
  Milky Way dark matter haloes: Aq-A, Aq-B, Aq-C, and Aq-D
  \citep{Springel08b}. These were run at Aquarius resolution level 3
  (softening length 120.5 pc, particle mass $
  5.6\times10^{4}, 2.5\times10^{4}, 5.4\times10^{4}$ and
  $5.4\times10^{4} \Msun$ respectively) with the \textsc{p-gadget3}
  code; the cosmological parameters were 7-year \emph{Wilkinson Microwave Anisotropy Probe} \citep[\emph{WMAP-7}][]{wmap11}. Haloes and subhaloes
  were identified using the gravitational potential unbinding code,
  \textsc{subfind} \citep{Springel01}. Spurious subhaloes -- those subhaloes that form by spurious fragmentation of filaments -- were
  identified and removed from the catalogues using the Lagrangian
  region shape and maximum mass criteria of \citet{Lovell14}. We then compare the
  conditional mass functions of these simulations with those derived
  from the EPS method. For a halo of mass $M_2$ at $z_2=0$, the
  conditional mass function gives the fraction of its mass is
  contained within progenitor haloes of mass $M_1$ at some earlier
  redshift $z_1$. 
  
  \begin{figure}
  	\includegraphics[scale=0.68]{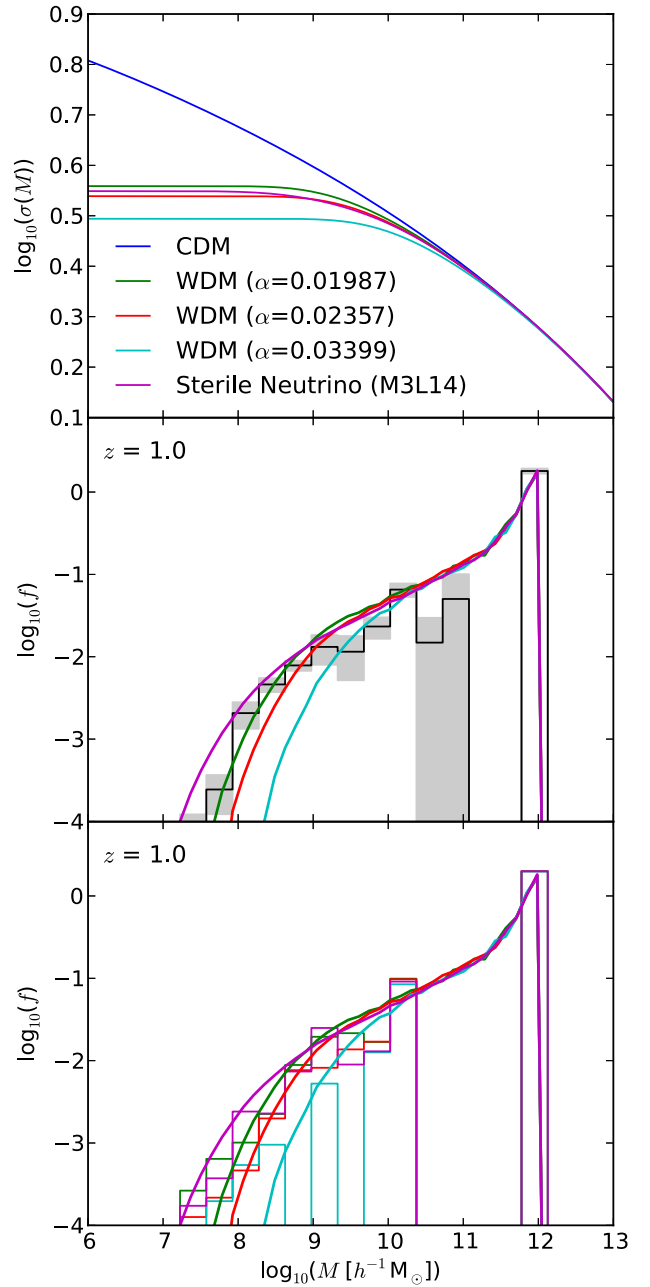}
        \caption{Top panel: $\sigma(M)$, the rms density fluctuation smoothed over mass scale, $M$, using a sharp $k$-space filter with $a=2.7$, for CDM, WDM and sterile neutrinos where, for WDM, $\alpha$ determines the position of the cutoff in the power spectrum. Middle panel: mean conditional mass function of four \emph{N}-body sterile neutrino haloes, based on the Aquarius Project with \emph{WMAP}-7 cosmological parameters (black histogram); $1\sigma$ errors are shown by the grey shaded area. Colour lines are the mean conditional mass functions of 1000 Monte Carlo simulations, using a sharp $k$-space filter, for the different DM cases, with the same colours as in the top panel. Bottom panel: \emph{N}-body conditional mass functions from halo A for the different DM cases (colour histograms). Curved lines are the same Monte Carlo conditional mass functions from the middle panel.}
  	\label{EPScomp}	
  \end{figure}
  
  We plot the conditional mass functions at $z_1=1$ in
  Fig.~\ref{EPScomp}, where the haloes have a final mass of
  $M_2\sim1.5\times10^{12}h^{-1}\Msun$. In the top panel we compare the rms density fluctuations of the
  M3L14 matter power spectrum with those of CDM and also three
  V05 thermal relic power spectra with transfer function
  parameters $\alpha=$0.0199, 0.0236 and 0.0340 $h^{-1}\rmn{Mpc}$,
  which correspond to thermal relic masses $M_{\rmn{th}}=$ 2.3, 2.0,
  and 1.5~~keV respectively \citep{Lovell14}. These were calculated
  using a sharp $k$-space filter with $a=2.7$. The M3L14 model has a
  different behaviour than the thermal relic models, in
  that the curve peels away from CDM at the same mass scale as the
  $\alpha=0.0236$ model but has a slightly shallower slope for large masses. Compared
  to the $\alpha=0.0236$ thermal relic $\sigma$ has a lower amplitude 
  at intermediate mass scales but a higher amplitude for
  $M<10^{9}h^{-1}\Msun$. This change is reflected in the $z_1=1$ conditional
  mass functions: $M\gsim10^{9}h^{-1}\Msun$, M3L14 produces a similar number of haloes as the
  $\alpha=0.0236$ thermal relic model, but below this mass the rate of
  decrease is much shallower such that at $M\sim10^{8}h^{-1}\Msun$
  M3L14 has a greater abundance of haloes than even the
  $\alpha=0.0199$ thermal relic model. In spite of this change, there is still
  good agreement between the number of substructures predicted by the EPS method
  and the number measured in the cleaned simulation halo catalogues. We
  choose a value of $a=2.7$ as this produces the best agreement, but
  the effect of varying $a$ is small.
  
  \subsection{Comparing the number of semi-analytic satellites to observations}
  
  The procedure for determining the number of satellites required for
  a model to be acceptable is the same as in \citet{Kennedy14}. We retain from that work the conservative assumptions that the census of classical satellites is complete across the entire sky (11 satellites including the Magellanic Clouds), and that the distribution of likely Milky Way satellites from more recent surveys is isotropic.

  
  This last assumption is complicated by the recent discovery of $\sim13$ satellites by the Dark Energy Survey\citep[DES][]{Bechtol15,DrlicaWagner15,Koposov15}, whose identity as satellites rather than globular clusters is established from their half-light radii, $r_\rmn{h}$, all of which exceed 30~pc. Up to 11 of those DES satellites with $M_V<-2$ are likely to have been satellites of the Large Magellanic Cloud \citep[LMC][]{Jethwa16}. Regardless of their origin, the satellites have been shown in these studies to be anisotropic in their distribution across the DES footprint. In order to incorporate these satellites into our algorithm, we add the 11 likely LMC satellites to the list of classical satellites which we assume is complete across the sky, and add the remaining two DES satellites -- along with the corresponding extra sky coverage --  to the `isotropic' Data Release 5 (DR5) selection that also includes the 11 Sloan Digital Sky Survey (SDSS) DR5 satellites. This choice enables us to take account of the LMC-Small Magellanic Cloud (SMC) system without almost certainly overestimating the number of unobserved satellites. In addition, analyses of the VST-Atlas and Pan-STARRS $3\pi$ surveys have discovered one \citep{Torrealba16} and three \citep{Laevens15a, Laevens15b} new satellites respectively. The completeness of these two surveys to date is very uncertain and so to include these satellites in our analysis we make the conservative assumption that they form part of the population assumed to be complete across the sky.
  
  Since SDSS DR5 observed 19 per cent of the sky, and the DES 2-yr study an additional 10 per cent, we are therefore interested in the number of `isotropic distribution satellites' within 30 per cent of the sky. For each galaxy generated by the semi-analytic model we determine the number of satellite galaxies within the halo virial radius having $M_{V}<-2$, and compare this to the number of Milky Way satellites within this same virial radius inferred from the observations.  We then iteratively determine the minimum mass\footnote{The masses returned by \textsc{galform} are defined as `Dhalo' masses, which constitute the gravitationally bound mass of the halo. For a given relaxed halo the Dhalo mass is typically less than 20 per cent larger than the measured virial mass, $M_{200}$ \citep{Jiang14}. Therefore when quoting values for \Mh we will assume they are equivalent to $M_{200}$.} of the host halo that would produce the required number of satellites for a given sterile neutrino model, which we define formally to be the smallest halo mass that, for 200 merger tree realizations, produces enough satellites at least 5 per cent of the time. We denote this minimum halo mass as \Mh; full details of the procedure can be found in \citet{Kennedy14} .

  \section{Results}
  \label{sec:res}
 
\subsection{Constraints on sterile neutrino parameters}
 
\begin{figure}
   \includegraphics[scale=0.45]{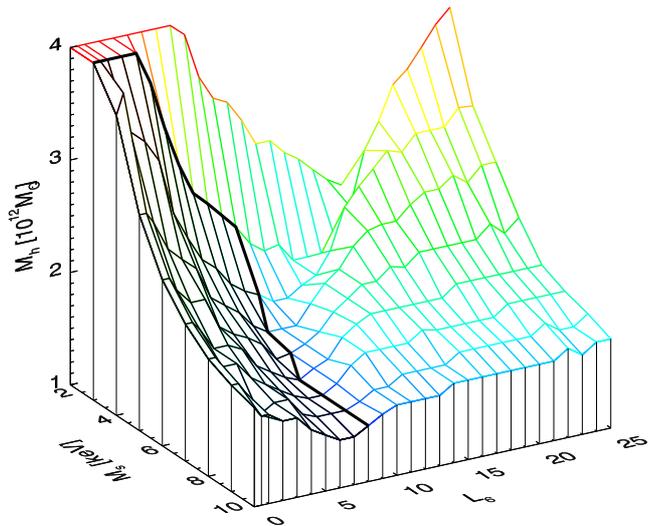}
   \caption{The surface that shows the values of \Mh as a function of sterile neutrino mass \Ms and lepton asymmetry $L_{6}$ that are consistent with the Milky Way satellite data. The plot is cropped such that the maximum permitted value of $\Mh=4\times10^{12}\Msun$. The set of models that are excluded at at least 95 per cent confidence by the X-ray limits of \citet{Watson:11} and \citet{Horiuchi:13} are shaded in black.} 
   \label{MH3D}
 \end{figure}

 We begin the presentation of our results with a 3D plot of \Mh as a function of both sterile neutrino mass and lepton asymmetry in Fig.~\ref{MH3D}. The range of sterile neutrino masses is 2-10~keV, and of $L_6$ is 0-25. We also compare the results with limits on \Ms-$L_6$ from X-ray decay non-detections \citep[95 per cent confidence limit:][]{Watson:11,Horiuchi:13}. 
  
 The allowed values of \Mh decrease as the sterile neutrino mass increases, in the same manner as one would expect for a thermal relic. The non-monotonic behaviour with lepton asymmetry is reflected in the fact that, for a given \Ms, \Mh attains a minimum value for a specific value of $L_{6}$, which we denote $L_{6,\rmn{min}}$. The value of $L_{6,\rmn{min}}$ decreases with sterile neutrino mass, falling from   $L_{6,\rmn{min}}=16$ at \Ms=3~keV to  $L_{6,\rmn{min}}=6$ at \Ms=10~keV, producing a characteristic winding valley shape in the surface. For \Ms$>$5~keV the position of the valley floor happens to coincide with the constraints from X-ray observations. Thus, further non-detections in this mass range would force the sterile neutrinos to be `warmer'. We also ran the same set of models for the fiducial value of $\alpha_{\rmn{hot}}$ -- 3.2 -- and found an increase of no more than 10 per cent in \Mh for any of our \Ms -- $L_{6}$ combinations. Thus, we are confident that the choice of $\alpha_{\rmn{hot}}$ calibration makes little difference to the results.
  
\begin{figure}
   \includegraphics[scale=0.35]{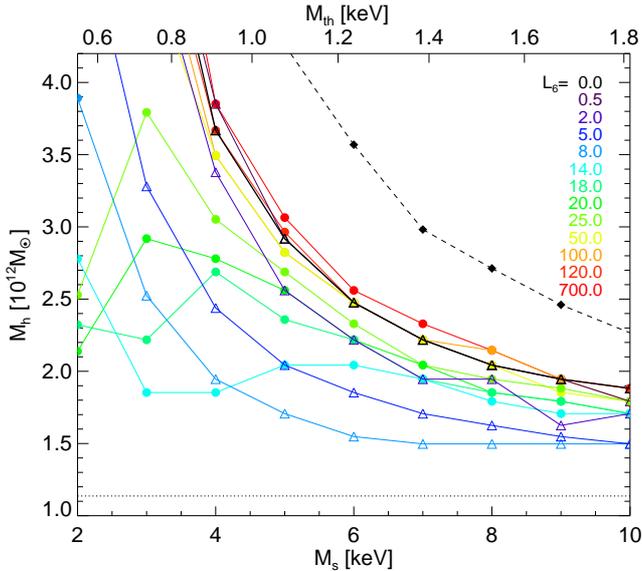}
   \caption{The Milky Way halo mass required to account for the observed number of Milky Way satellites, plotted as a function of sterile neutrino mass. Regions above and to the right of a given line are allowed; those below and to the left are disallowed. Colour lines show constraints for different values of the lepton asymmetry. Empty triangles denote \Ms-$L_6$ combinations that are excluded by X-ray non-detections (black region in Fig.~\protect\ref{MH3D}), as presented in \citet{Boyarsky14a}, and filled circles those that are not excluded. The dashed black line is the constraint for trees generated with the V05 thermal relic fitting function discussed in Section~\ref{sec:snm}. The thermal relic masses, calculated from equation~\ref{SterTherm}, are shown on the top axis. The dotted line shows the result obtained for CDM.}
   \label{MHMall}
 \end{figure}

 We expand on our results in Fig.~\ref{MHMall}, in which we plot a separate \Mh-\Ms relation for each value of $L_{6}$. We have also generated merger trees for thermal relic power spectra using  the V05 transfer function for  thermal relic masses in the range [0.79-1.8]~keV.  
  
 \begin{figure}
   \centering
   \includegraphics[scale=0.38,angle=-90]{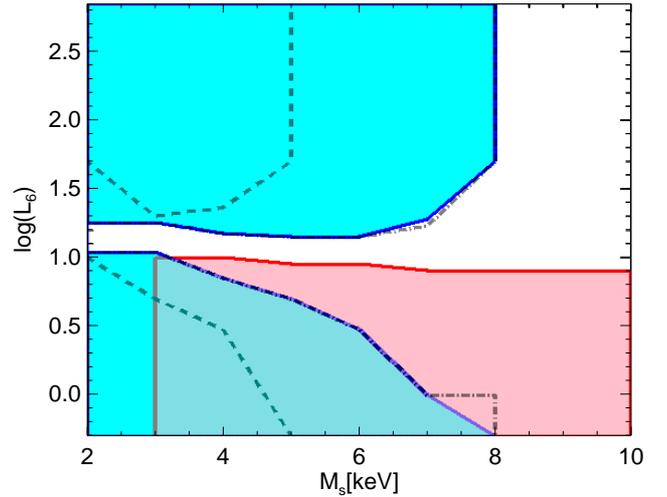}
   \caption{Excluded region in the $\Ms$-$L_6$ plane. For sterile neutrino parameters in the cyan shaded region the correct number of Milky Way satellite would require the mass of the halo, $M_{\rm h} > 2\times 10^{12}\Msun$, which is disfavoured by current astronomical data. The red shaded region is excluded from a non-observation of the X-ray decay line. Even for masses as low as $\Ms =2$~keV there exist primordial distribution functions ``cold enough'' to predict the Galactic structures in accordance with observations. The dashed and dashed-dotted lines
show how the excluded regions change when one of the \textsc{galform} parameters is varied.
The dashed line shows the effect of change of
$v_{\rm cut}$ from its fiducial value ($30$~\kms) down to $v_{\rm cut} = 25~\kms$. The dashed-dotted line shows the effect of change of $z_{\rm cut}$, from its default value $z_{\rm cut} = 10$ to $12$; see the text for parameter definitions.}
   \label{fig:excluded_Ms_L6}
 \end{figure}

Spectra with $L_{6}=0$ and $700$ result in very similar allowed values of \Mh for a given sterile neutrino mass, as expected from their input matter power spectra. The minimum acceptable value of $M_\rmn{h}$ drops by over a factor of 2 between the most extreme values of $L_6$ and the optimal value $L_6\sim10$, especially at the lowest masses where the results are predominantly determined by the sterile neutrino transfer function rather than by the galaxy formation physics. However, the lowest possible value of \Mh is given by CDM, which occurs at \Mh$=1.16\times10^{12}\Msun$: therefore the sterile neutrino models with still higher masses may return a value of \Mh up to $\sim3\times10^{11}\Msun$ lower before hitting the CDM limit.

 We can seek to rule out models by making conservative assumptions about the mass of the Milky Way halo. Estimates of the Milky Way halo mass have been made using a variety of methods, including  the Local Hubble Flow \citep{Penarrubia14}, dynamical tracers \citep{Deason12, Piffl14,Wang15}, the kinematics of bright Milky Way satellites, \citep{Sales07a,Sales07b,Busha11,Wang12,BK13,Gonzalez13,Cautun14b} and the timing argument \citep{Kahn59,Li08}; for a more comprehensive list see \citet{Wang15}. Together these studies allow a wide range of possible halo masses, 
 
 \begin{equation}
0.5\times10^{12} <M_{200}<2\times10^{12}\Msun\label{eq:1}
\end{equation}

where $M_{200}$ is the mass enclosed within the radius of overdensity 200. In order to set a conservative constraint, \emph{we take our upper limit to be $\Mh=2\times10^{12}\Msun$}; here we have again assumed that the Dhalo definition of mass in our model is approximately equal to $M_{200}$. This value rules out a swathe of models with \Ms$<7\rmn{~keV}$. We find that the non-resonant sterile neutrino ($L_{6}=0$) is ruled out for \Ms$\leq 8~\rmn{keV}$. This bound is weaker, although compatible with the SDSS-based Lyman~$\alpha$ bounds \citep{Seljak:06,Viel:06,Boyarsky:08c}; it is comparable with the other structure formation bounds~\citep{Maccio:09,Polisensky2011,Schneider13,Lovell14}.

However, for the resonantly produced sterile neutrinos there exists a range of $L_{6}$ which is compatible with the entire allowed range of halo masses (see Fig.~\ref{fig:excluded_Ms_L6}). In fact, all sterile neutrino masses $>3$~keV are permitted for $L_{6}\sim14$ if the \Mh constraints alone are considered. When combined with the limits from X-ray non-detections, the approximate constraint on sterile neutrino lepton asymmetry in the mass range $3\le\Ms/\rmn{keV}<6$ is $10\lsim L_6\lsim16$. The upper limit weakens substantially for 7~keV sterile neutrinos, and disappears for $\Ms>9$~keV.


We have previously shown that our results are sensitive to the $\alpha_{\rmn{hot}}$ feedback parameter at the 10 per cent level, however \citet{Kennedy14} showed that there exist other systematic uncertainties within the galaxy formation model that may have a larger effect, and in particular the parameters regulating reionization feedback. {\sc galform} models reionization such that haloes in the merger tree that have a circular velocity lower than some parameter $v_{\rmn{cut}}$ at a redshift of $z_{\rmn{cut}}$ do not contain any gas: this reflects the heating of the gas by a background of ionizing photons \citep{Cole00,Benson06}, thus preventing it from ever falling into the halo. In the \citet{Gonzalez14} version of {\sc galform} $v_{\rmn{cut}}=30~\kms$ and $z_{\rmn{cut}}=10$. In Fig.~\ref{fig:excluded_Ms_L6} we also derived limits for $z_{\rmn{cut}}=12$ and $v_{\rmn{cut}}=25~\kms$. The former has very little effect on our results, however the lowering of $v_{\rmn{cut}}$ introduces many more small galaxies and thus weakens our limits substantially. An increase in  $v_{\rmn{cut}}$ would have the exact opposite effect.  
  
\begin{figure}   
 	 \includegraphics[width=\linewidth]{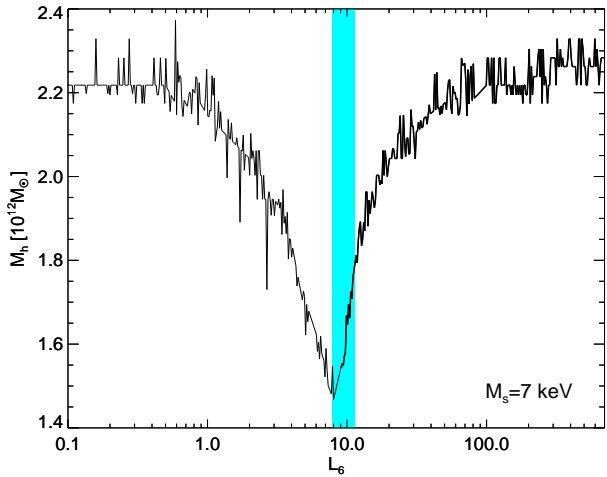}
    \caption{Minimum Milky Way halo mass, \Mh, as a function of $L_6$ for
    $\Ms=7$~keV. The region in $L_6$ consistent with the 3.5~keV line is shown in cyan. Data points that are in tension with the X-ray constraints \protect\citep[$L_6<9$;][]{Watson:11,Horiuchi:13} are connected by thin lines, the remainder by thick lines. }
    \label{MHM7}
  \end{figure}

 In order to explore the behaviour of the model in greater detail, and determine precisely the consequences for the $3.5$~keV line/7~keV sterile neutrino, we calculate \Mh for the fine grid of $\Ms=7$~keV models used in Figs~\ref{L6vsQav}~and~\ref{L6vsPG}. The results are plotted in Fig.~\ref{MHM7}.  
  
 The profile of allowed values of \Mh again has a valley shape, and sharp features are introduced by stochastic scatter in the number of halo satellites. The lowest allowed value of \Mh~ is $1.5\times10^{12}~\Msun$, which is comfortably within the range estimate for the Milky Way. This occurs at $L_6\approx8$, and either side of this, the constraint on \Mh rises rapidly to a maximum of $\approx2.3\times10^{12}~\Msun$. The region consistent with the 3.5~keV line falls near the centre of the trough in \Mh. Thus, the minimum allowed \Mh for the line is $1.5\times10^{12}\Msun$. As was the case for the full range of \Ms, the X-ray non-detection 95 per cent exclusion range terminates at the bottom of the trough, such that further, more stringent non-detections would force \Mh to increase. It is also the case that if the `true' momentum distribution function exhibits a stronger back reaction than is assumed here, \Mh would increase and the limits become stronger. However, for the present study we are content that values of lepton asymmetry in the range $2\lsim L_6\lsim 20$ are consistent with the current estimate of the Milky Way halo mass.
 
\subsection{Comparison to thermal relics}
 
We now compare these results for sterile neutrinos with those for V05 thermal relics as used in \citep{Kennedy14}. They restricted their analysis to relics of mass 1.5~keV and larger; we extend our analysis to lower masses to show the trend obtained, even though it would be unlikely that we could recalibrate the model to get the correct luminosity functions. We therefore compute nine V05 thermal relic power spectra in the mass range [0.7,1.8]keV and perform the same semi-analytic procedure as used on the sterile neutrino models. These are included in Fig~\ref{MHMall}, with the equation~\ref{SterTherm} conversion from sterile neutrino mass to thermal relic mass shown along the top $x$-axis.
 
 We find that sterile neutrino power spectra with V05 thermal-relic equivalent masses as low as 1.2~keV\footnote{ This result, along with those in the rest of this subsection, has been derived using the SDSS satellites alone in order to make a fair comparison with previous work. When the DES satellites are included, as shown in Fig.~\ref{MHMall}, this miniumum thermal relic mass rises to $>1.8$~keV.} can generate enough satellites and be compatible with our adopted upper limit of $\Mh<2\times10^{12}\Msun$, whereas previous studies that used the V05 fit in simulations have shown that the thermal relic mass should be at the very least $1.6$~keV \citep{Lovell14} and preferably higher still \citep[>2.3~keV][]{Polisensky2011}. \citet{Kennedy14} found that a 1.5~keV thermal relic generated from V05 would require $\Mh\approx2.2\times10^{12}\Msun$: our V05 relic calculation returns $\Mh\approx2.3\times10^{12}\Msun$, which is encouraging given that the choice of  {\sc galform} model and cosmology differ between their study and ours. When instead considering our results that include the DES satellites, the \Mh results for the V05 relics are $\approx25$ per cent higher than would be obtained for our sterile neutrinos with the corresponding V05 thermal relic mass. Therefore care should be taken when comparing our results to previous studies.

  \section{Conclusions}
  \label{sec:con}

  Sterile neutrinos are an intriguing dark matter
  candidate. The underlying model of particle physics can
  potentially solve many unanswered questions in astronomy and particle
  physics, and provides a possible explanation for the unidentified 3.5~keV
  X-ray line  recently seen in galaxies and clusters \citep{Boyarsky14a,Boyarsky14b,Bulbul14}. In addition to its role as an X-ray emitter, the
  sterile neutrino can act like warm dark matter and have properties
  that are imprinted on the abundance and structure of the Milky Way's
  satellite galaxies. We therefore set out to calculate how the parameters of the sterile neutrino model can be constrained by requiring that the expected number of satellite galaxies in the Milky Way for our chosen galaxy formation model should match the observed number of Milky Way satellites.
  
  We reviewed the history of sterile neutrinos as a dark matter candidate, beginning with the initial suggestion of a single, non-resonantly produced sterile neutrino \citep{Dodelson94}. This new particle was motivated by its ability to generate neutrino oscillations and to make up the dark matter. However it was later shown to be incompatible with non-detections of X-ray decay radiation at large masses and with structure formation constraints at lower masses, such that the model was ruled out \citep{Seljak:06,Viel:06}. Later work showed that the presence of a lepton asymmetry could alleviate both of these constraints \citep{Shi99}. It could allow for a smaller neutrino mixing angle, weakening the X-ray bounds, and through resonant production enhance the production of low-momenta sterile neutrinos, thus resulting in better agreement with the observed properties of galaxies. A series of studies then found that the necessary lepton asymmetry could be generated as part of a wider theory that included two further sterile neutrinos at the GeV scale~\citep[e.g.][]{Asaka:05b}. The theory as a whole could then explain neutrino oscillations, dark matter, and baryogenesis. When coupled to the possible detection of an X-ray decay line \citep{Boyarsky14a,Boyarsky14b,Bulbul14}, the keV sterile neutrino had become a competitive and compelling dark matter candidate.
  
  We discussed how the properties of sterile neutrino dark matter are very similar to those of warm dark matter, in that the sterile neutrinos are able to freestream out of small perturbations in the early Universe, thus truncating the matter power spectrum and preventing small galaxies from forming. We showed how the presence of the lepton asymmetry alters the sterile neutrino momentum distribution function, and how the presence of resonances enhances the number of low momentum particles and thus produce a cooler matter power spectrum. The effectiveness of the lepton asymmetry to produce an arbitrarily cold power spectrum was, however, seen to have its limits, in that large lepton asymmetries enhance production at all momenta and result in a very similar power spectrum to that obtained for standard warm dark matter. We also compared the sterile neutrino matter power spectra to those described by the fits used in Lyman-$\alpha$ studies, and found that the latter underestimates the power on comoving scales $>5~h~\rmn{Mpc}^{-1}$ due to subsequent developments in the calculation of sterile neutrino momentum distributions.
  
 The sterile neutrino model described above is a two parameter model, with a particle mass, \Ms, and a value of the lepton asymmetry, $L_6$. We assembled a grid of \Ms and $L_6$ pairs, generated their matter power spectra, and used these spectra to build dark matter halo merger trees. We then populated these haloes with galaxies using the {\sc galform} semi-analytic model of galaxy formation and evolution, and thus calculated their satellite galaxy abundances as a function of the assumed Milky Way halo mass. Thus the number of satellites depends both on the sterile neutrino properties and on the Milky Way halo mass. We found that heavier sterile neutrinos are viable candidates even if the Milky Way halo mass is at the lower end of its observationally allowed range, as is the case for thermal relics. For each particle mass we can associate a characteristic lepton asymmetry associated with it for which the allowed lower limit on the halo mass is minimized. By assuming that the Milky Way halo mass must be no higher than $2\times10^{12}~\Msun$, we were able to constrain the lepton asymmetry for sterile neutrino masses lower than 9~keV. We showed that the range of  7~keV lepton asymmetries favoured by the 3.5~keV X-ray line requires the Milky Way halo mass to be no less than $1.5\times10^{12}\Msun$, well within acceptable bounds. The relationship between minimum allowed halo mass and the wavenumber of the peak of the input matter power spectrum was shown to be tight, and any variation between models with the same peak location could be at least partly explained by differences in the power spectrum slope.    
 
 A major uncertainty in our analysis is the total number of satellites that orbit the Milky Way. This is uncertain because current surveys do not cover the whole sky and it is well known that at least the brightest satellites appear in an anisotropic configuration. The number of satellites already known to exist is already approaching the total number of subhaloes predicted to exist in even the coldest of the sterile neutrino models that are consistent with the 3.5keV line. A complete census of the satellites population of the Milky Way, which may be obtained with future surveys, may well conclusively rule out the 7~keV resonantly-produced sterile neutrino as the dominant component of the dark matter.
 
%
  Sterile neutrinos are thus a very interesting dark matter candidate, and the 7~keV candidate that could be the source of the 3.5~keV line is in good agreement with current observations of the Milky Way satellites. There remain nevertheless many uncertainties in the argument presented in this paper. The main limitations are uncertainties in the galaxy formation model, particularly the treatment of supernova and photoreionization feedback and how these affect the mass of their host haloes, although the current generation of simulations \citep{Sawala14,Fattahi15,Onorbe15} are providing valuable insights, at least for CDM. Observationally, the total number and radial distribution of satellites remains poorly constrained. An exciting and powerful development will be X-ray constraints from other observational targets. In particular, observations of dark matter-dominated satellites have the potential to either rule out much more of the $M_{\rmn{s}}-L_6$ parameter space or to obtain further detections of the 3.5~keV line. Such a detection could be used in tandem with dwarf spheroidal mass estimates \citep{Walker09,Walker10,Wolf10} to determine the sterile neutrino mixing angle and therefore the lepton asymmetry. This could lead to an accurate determination of the matter power spectrum, and thus perhaps to a new paradigm in galaxy formation and cosmology.


\section*{Acknowledgements}

MRL would like to thank  Keith Bechtol, Artem Ivashko and Antonella Garzilli for useful discussions. We would like to thank Mikko Laine for supplying the code that calculates the sterile neutrino distribution functions. This work used the DiRAC Data Centric system at Durham University, operated by the Institute for Computational Cosmology on behalf of the STFC DiRAC HPC Facility (www.dirac.ac.uk). This equipment was funded by BIS National E-infrastructure cap- ital grant ST/K00042X/1, STFC capital grant ST/H008519/1, and STFC DiRAC Operations grant ST/K003267/1 and Durham University. DiRAC is part of the National E-Infrastructure. This work is part of the D-ITP consortium, a programme of the Netherlands Organization for Scientific Research (NWO) that is funded by the Dutch Ministry of Education, Culture and Science (OCW).  This work was supported in part by an STFC
 rolling grant to the ICC and by ERC Advanced Investigator grant COSMIWAY [GA 267291]. SC and AS acknowledge support from STFC grant ST/L00075X/1. SB is supported by STFC through grant ST/K501979/1. VGP acknowledges support from a European Research Council Starting Grant (DEGAS-259586)

  \bibliographystyle{mnras}

\bsp
  \label{lastpage}
  
  \appendix
  \section{Feedback calibration}
\label{AFC}	
  
  It is a well known issue in astrophysics that stellar feedback plays a crucial role in regulating the stellar mass of galaxies in general, and thus the number of bright Milky Way satellites in particular. In the model of \citet{Gonzalez14} the strength of stellar feedback is parametrized in terms of the mass loading parameter, $\beta$, which in this model is defined as the ratio of the gas ejection rate and the star formation rate.  $\beta$ is a function of the circular velocity of the disc and the bulge, $v_{circ}$, and takes the form:
  \begin{equation}
    \beta = \left(\frac{v_{\rmn{circ}}}{v_{\rmn{hot}}}\right)^{-\alpha_{\rmn{hot}}},	
  \end{equation}
where $v_{\rmn{hot}}$ and $\alpha_{\rmn{hot}}$ are parameters of the model; their fiducial values are deemed to be those that provide a good fit to the local $b_{J}$- and $K$-band luminosity functions, and in this model take the values $v_{\rmn{hot}}=425$~\kms  and $\alpha_{\rmn{hot}}=3.2$. 
  
\begin{figure}
  	\includegraphics[scale=0.34]{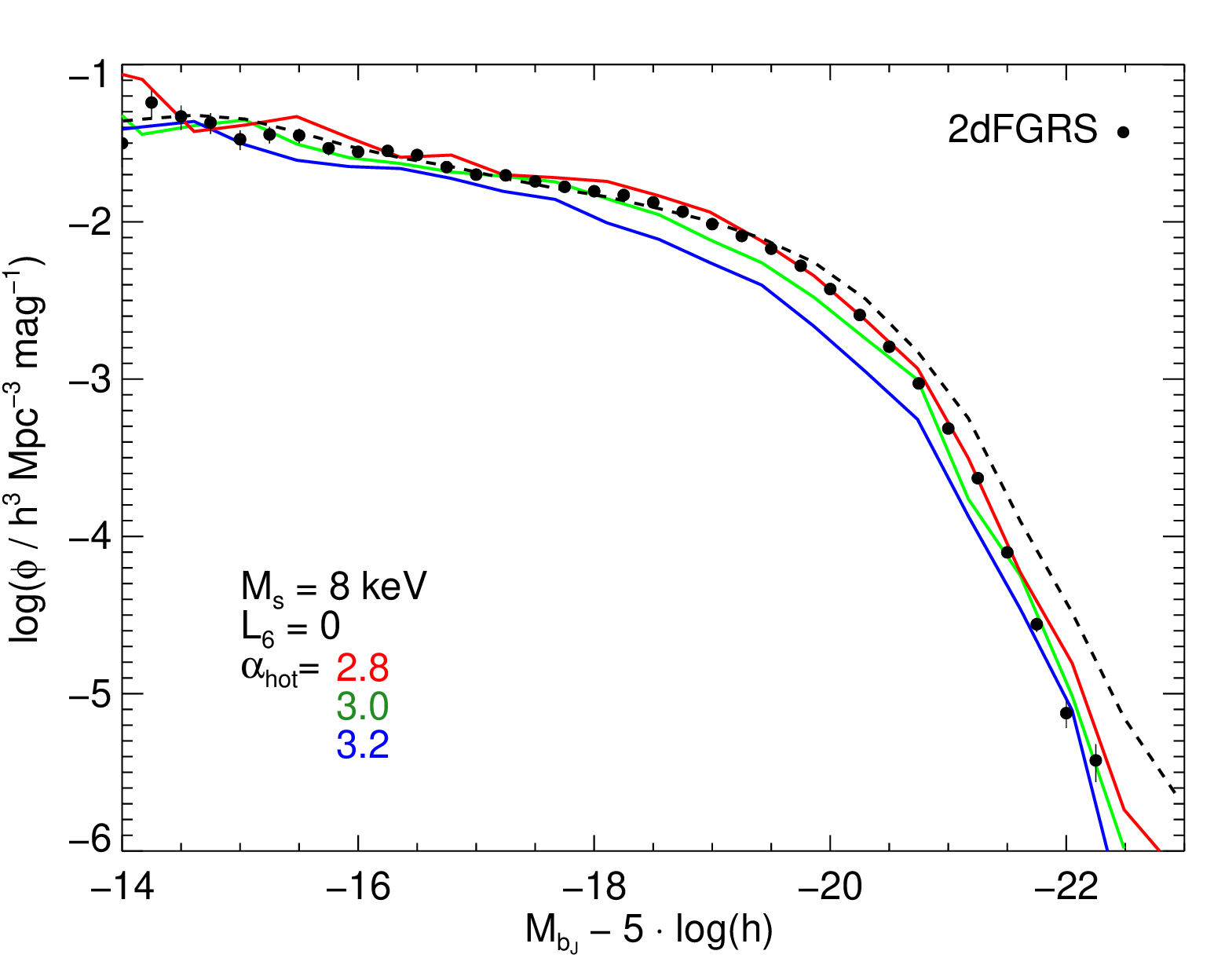}
	\caption{The $b_J$-band luminosity function at $z=0$ for an 8~keV sterile neutrino (NRP) with $\alpha_{\rmn{hot}}=$2.8, 3.0, and 3.2 displayed as red, blue, and green lines respectively. The observational data from the 2dFGRS \citep{cole05} are shown as filled circles. The fiducial CDM model as developed in \citet{Gonzalez14} is shown as a dashed black line.}
	\label{bJLF}
  \end{figure}

  \citet{Kennedy14} showed that the lower concentrations of  WDM haloes result in lower values of $v_{\rmn{circ}}$ for a halo of a given mass, and that therefore a recalibration is required to still match the observed $b_{J}$- and $K$-band luminosity functions at $z=0$: they required $\alpha_{\rmn{hot}}=3.0$ for the 1.5~keV thermal relic and proposed a functional form to fit all cooler models up to the limit of $\alpha_{\rmn{hot}}=3.2$ for CDM. We extend this result to one of our sterile neutrino models in Fig.~\ref{bJLF}, in which we display the $b_{J}$ luminosity function for a sterile neutrino with $M_{\rmn{s}}=8$~keV and $L_6=0$ (i.e. NRP) when adopting $\alpha_{\rmn{hot}}=2.8, 3.0$, and 3.2. We select 8~keV NRP because it is expected to have a power spectrum similar to that of the 1.5~keV thermal relic limit considered by \citet{Kennedy14}; 1.5~keV is also the 99.7 per cent confidence limit lower bound from Lyman-$\alpha$ constraints inferred by \citet{Boyarsky:08d}. The fiducial value --i.e. that used for CDM -- of $\alpha_{\rmn{hot}}=3.2$  underpredicts the number of galaxies across all magnitudes as expected. The other two values of $\alpha_{\rmn{hot}}$ lie much closer to the observations, with $\alpha_{\rmn{hot}}=3.0$ underpredicting the number of galaxies at magnitudes $\sim-19$ and $\alpha_{\rmn{hot}}=2.8$ slightly overpredicting the number of both very luminous and faint galaxies. It is therefore necessary to make an adjustment to the fiducial model to account for the differences in WDM structure formation. 
  
  For the present study one approach would be to recalibrate $\alpha_{\rmn{hot}}$ for all of our transfer functions individually. Another option would be instead to adopt the value fitted by \citet{Kennedy14}, $\alpha_{\rmn{hot}}=3.0$, and apply this to all of our models. In this case the models will overpredict the number of satellites for those combinations of $L_{6}$ and \Ms that are not already in severe tension with Lyman-$\alpha$ constraints; any combination that nevertheless fails to produce enough satellites for the expected range of halo masses will be ruled out confidently. We therefore adopt the conservative value $\alpha_{\rmn{hot}}=3.0$ for all the results published here. In several cases we have also performed runs with the fiducial value of $\alpha_{\rmn{hot}}=3.2$, and compared the two sets of results in the text.
  
   There are other relevant parameters that can have important effects on the satellite galaxy abundance, such as the minimum value of $v_{\rmn{circ}}$ required for a galaxy to retain its gas after reionisation, and have a brief discussion of this topic in section~\ref{sec:res}. However, the primary purpose of this study is to carry out an exploration of the kind of constraints that can be imposed on sterile neutrino models from the observed abundance of Milky Way satellites. We leave more detailed investigation into the effects of galaxy formation to future work. See \citet{Kennedy14} for a discussion of these parameters.
  
  \section{Relationship between model characteristic scales and minimum halo mass}   
  \label{app:char}
\begin{figure}
  	\includegraphics[scale=0.34]{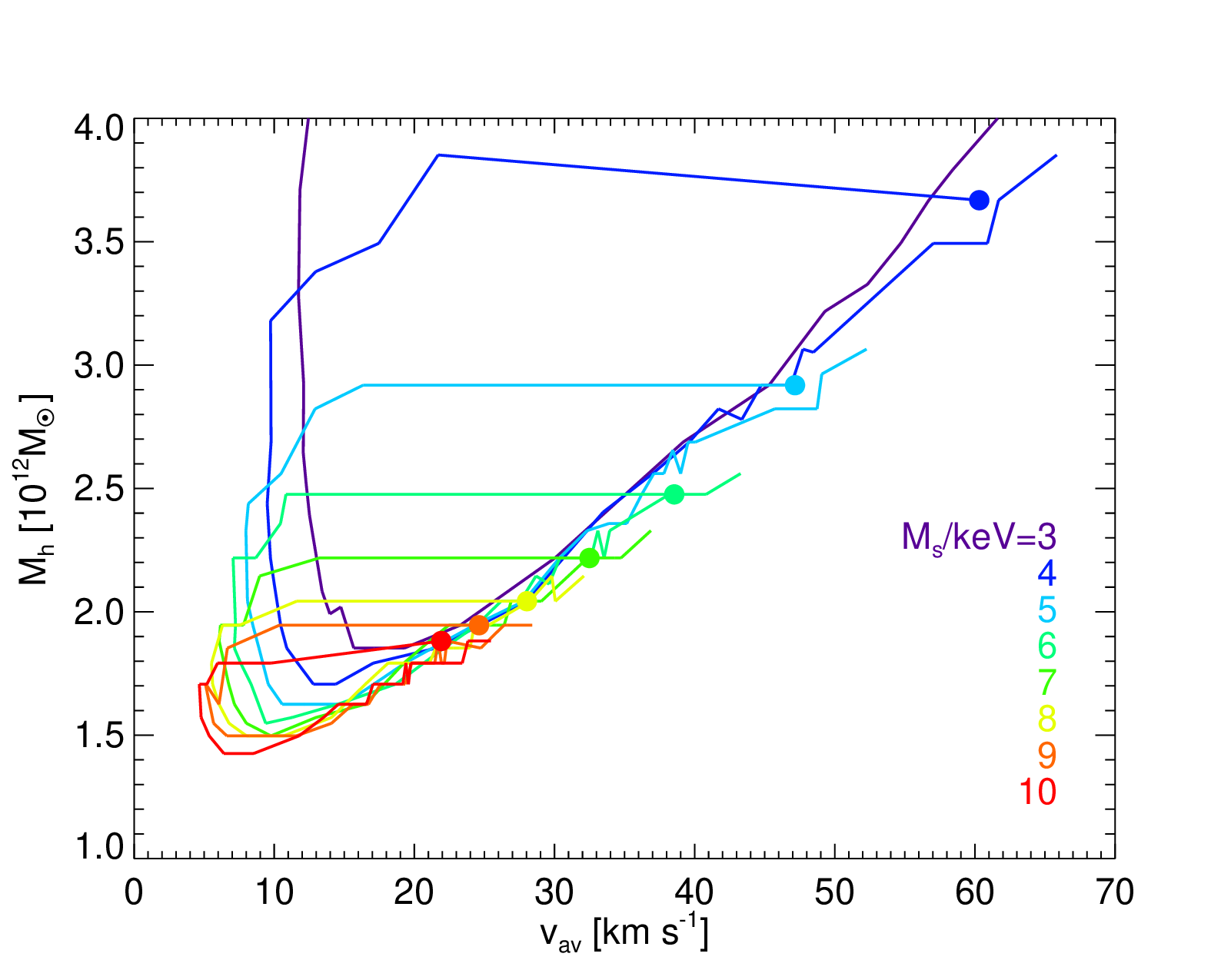}
	\caption{Minimum value of $M_{h}$ required to obtain the observed number of Milky Way satellites as a function of the average velocity, $v_{\rmn{av}}$, for \Ms in the range 3-10~keV. Each line corresponds to the series of $L_6$ for a given \Ms; the correspondence between line colour and \Ms is given in the legend. $L_6=0$ is denoted by a circle. Thick lines show regions that are in agreement with the X-ray non-detection bounds \citep[][]{Watson:11,Horiuchi:13} . }
	\label{VavMh}
  \end{figure}

  Having established the dependence of the minimum allowed value of \Mh on the particle physics parameters, sterile neutrino mass and lepton asymmetry, we now examine how this relationship is mediated by properties measured from the distribution function and the matter power spectrum. The first variable that we consider is the average velocity at matter-radiation equality, $v_{\rmn{av}}$, which we adopt as a proxy for the free-streaming velocity. Larger free-streaming velocities dampen more small-scale power and thus require a larger \Mh. We show the dependence of \Mh on $v_{\rmn{av}}$ for our \Ms-$L_6$ sample in Fig.~\ref{VavMh}, and highlight NRP models with circles so as to discern the additional features that resonant production entails.
  
  We find that for NRP sterile neutrinos, \Mh increases exponentially with $v_{\rmn{av}}$. The reduction in $v_{\rmn{av}}$ that results from the lepton asymmetry causes \Mh to decrease very slowly at first, and then plummets to a minimum before recovering gradually back towards the NRP solution. The lowest value of $v_{\rmn{av}}$ does not produce the lowest \Mh, a result that could be inferred by noting that the maximum value of \kpeak and minimum value of $v_{\rmn{av}}$ occur at different values of $L_6$ (c.f. Figures~\ref{L6vsQav} and ~\ref{L6vsPG}a.) For some \Ms, two values of \Mh at a given $v_{\rmn{av}}$ can  differ by over a factor of 2. We therefore conclude that, in general, $v_{\rmn{av}}$ is not a good predictor of \Mh. However, if one restricts attention to models with $L_6>L_{6,\rmn{min}}$, we do obtain a clear correlation. Given that for many values of \Ms it is precisely this range of $L_6$ that is permitted by the X-ray decay constraints, with the help of the latest galaxy formation models it may indeed be possible to predict \Mh from $v_{\rmn{av}}$ for X-ray-viable models. 

\begin{figure}
  	\includegraphics[scale=0.34]{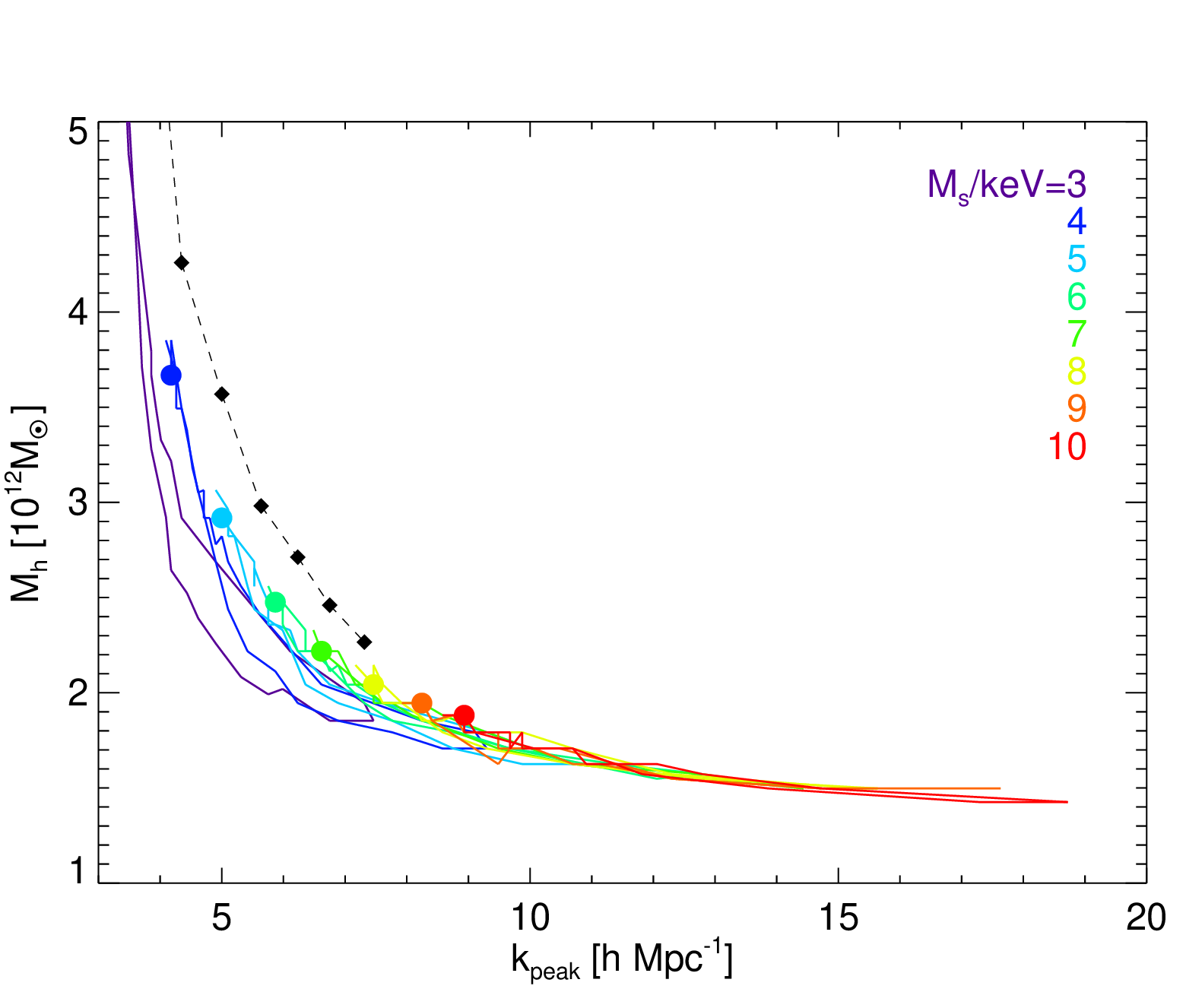}
	\caption{The minimum acceptable value of \Mh required to obtain the observed number of Milky Way satellites as a function of the matter power spectrum peak wavenumber, \kpeak, for \Ms in the range 3-10~keV.  The colours and symbols are the same as in Fig.~\ref{VavMh}. We also include the results for the V05 thermal relics, shown as diamonds.}
	\label{KpvMh}
  \end{figure}

An alternative property for determining \Mh is the power spectrum peak wavenumber, \kpeak. Larger values of this parameter allow more power on small scales, thus reducing \Mh. We take the values of \kpeak computed for each \Ms as a function of $L_6$ and plot the resulting \Mh--\kpeak relation in Fig.~\ref{KpvMh}.

  The dependence of \Mh on \kpeak is much cleaner than that on $v_{\rmn{av}}$.  \Mh falls rapidly with increasing \kpeak, with an apparent asymptote towards $1.4\times10^{12}~\Msun$ for $\kpeak~\to~\infty$ (i.e. CDM). The scatter in \Mh between all combinations of \Ms and $L_6$ that correspond to the same \kpeak is rarely more than 10 per cent, with the notable exception of the $\kpeak<5h~\rmn{Mpc}^{-1}$ regime and the V05 thermal relics. The latter have \Mh higher than sterile neutrinos at a given \kpeak by of order tens of per cent. This shows that it is not sufficient to define a generic WDM-like model purely by \kpeak; the shape of the cutoff is also important. 
  
\begin{figure}
  	\includegraphics[scale=0.34]{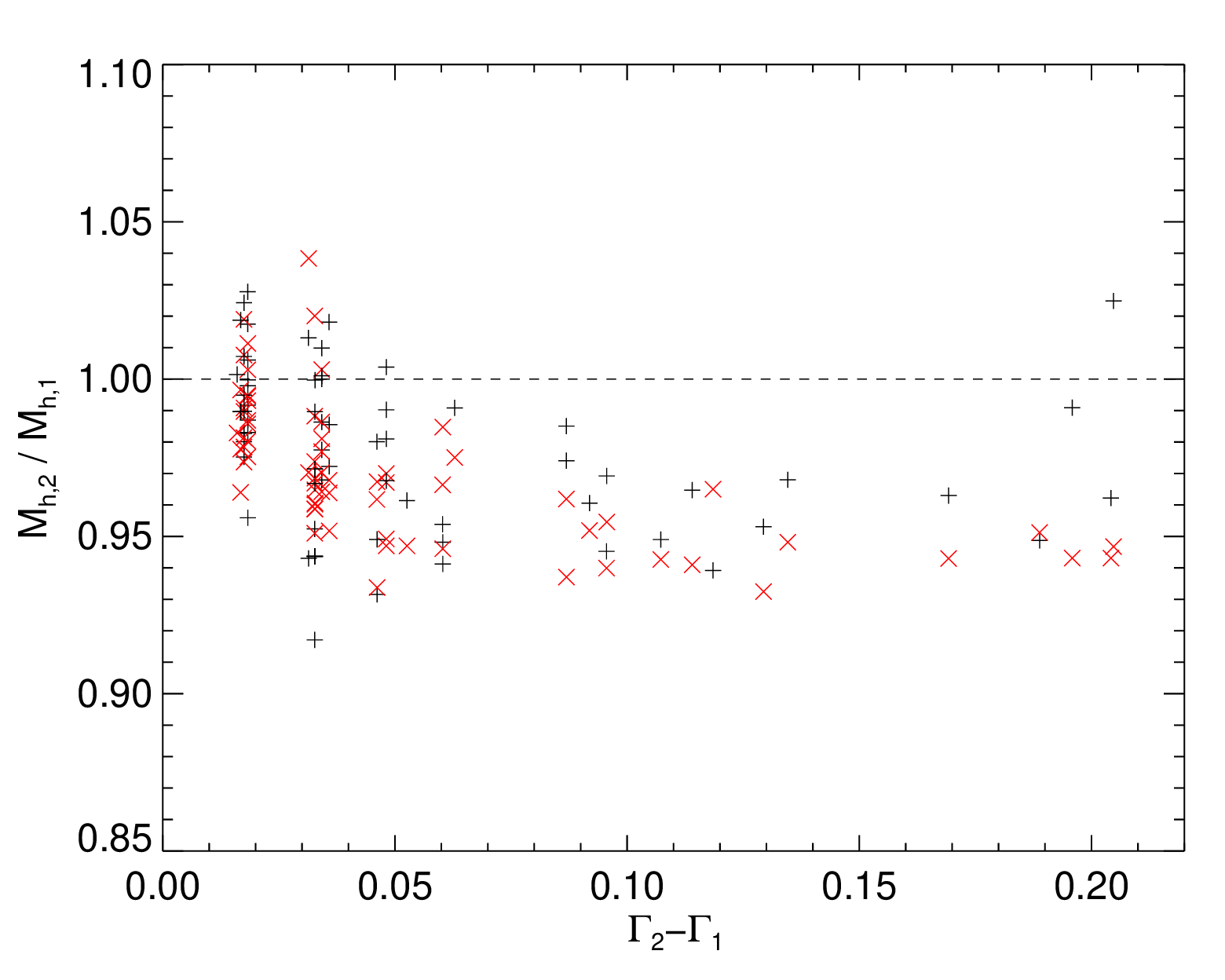}
	\caption{The ratio of \Mh as a function of the difference in $\Gamma$ for pairs of 7~keV models that share the same \kpeak. We plot raw data and smoothed data as black plus-signs and red crosses respectively.}
	\label{GamvMh}
  \end{figure}

  In order to explore the effect of the cutoff shape further, we consider the variation of the post-cutoff slope, $\Gamma$, with lepton asymmetry. We showed in Fig.~\ref{L6vsPG} that $k_\rmn{peak}$ and $\Gamma$ attain their maximum values at slightly different values of $L_6$: $k_\rmn{peak}$ is largest at $L_6=8$ whereas the peak value in $\Gamma$ occurs at $L_6\approx6.5$. This offset between the two has the consequence that for each value of \kpeak there are two possible values of $\Gamma$: a higher one for lower $L_6$ and vice versa. We investigate the effect of this property by finding pairs of $L_6$ in our sample of 500 7~keV models that have the same values of \kpeak, and calculate the difference between their two values of $\Gamma$ in addition to the ratio of their values of \Mh. Where the finite binning of \kpeak forces more than two models to take the same \kpeak, we consider the pair that have the largest difference in $\Gamma$. We display the results in Fig.~\ref{GamvMh}.
  
There is a trend in the raw data for \Mh to decrease for shallower slopes by, on average, 5 per cent for $\Gamma_2-\Gamma_1>0.05$. In order to reduce the noise due to halo-to-halo scatter, we smooth $\Mh(L_6)$ over the lepton asymmetry range $[\log{(L_6)}-0.077, \log{(L_6)}+0.077]$. The trend then exhibits much reduced scatter, such that all of the model pairs around $\Gamma_2-\Gamma_1>0.17$ exhibit a decrement of over 6 per cent. However, even the maximum decrement is too small to explain fully the tens of per cent discrepancy that is present for the V05 thermal relics, despite the similarity in the values of $\Gamma$ for V05 relics and NRP sterile neutrinos shown in Figure~\ref{L6vsPG}b. It may be that further variations in the spectrum shape not captured by our statistic $\Gamma$ are relevant for the production of satellites galaxies. Nevertheless, we have shown that the shallower slopes of some resonantly produced sterile neutrinos are the reason for the increase in the number of potential satellites, even in comparison to NRP models with the same \kpeak.

\end{document}